\pdfoutput=1

\documentclass[11pt]{article}
\usepackage{jheppub}

\usepackage{amsmath}
\usepackage{amssymb}
\usepackage{graphicx,color,slashed}
\usepackage{bbm} 
\usepackage{ifpdf}

\newcommand{\cF}{\mathcal{F}}

\newcommand{\cN}{\mathcal{N}}

\newcommand{\be}{\begin{equation}}
\newcommand{\ee}{\end{equation}}
\newcommand{\ba}{\begin{eqnarray}}
\newcommand{\ea}{\end{eqnarray}}
\newcommand{\nn}{\nonumber}
\newcommand{\lp}{\left(}
\newcommand{\rp}{\right)}
\newcommand{\ls}{\left[}
\newcommand{\rs}{\right]}

\newcommand{\w}{\wedge}

\newcommand{\N}{\mathcal{N}}

\def\rmi{{\rm i}}
\def\K{{K\"{a}hler}}
\def\ib{{\bar \imath}}
\def\jb{{\bar \jmath}}

\def\gold{{\zeta}}
\def\goldbar{{\bar\zeta}}
\def\tgold{{\tilde \gold}}
\newcommand{\kap}{{}}

\newcommand{\fermX}{{\psi^0}}
\newcommand{\fermXbar}{{\bar \psi^0}}

\newcommand{\ferm}[1]{{\psi^{#1}}}
\newcommand{\fermbar}[1]{{\bar \psi^{#1}}}
\newcommand{\fermb}[1]{{\chi^{#1}}}

\newcommand{\fermXb}{{\chi^0}}
\newcommand{\fermXbbar}{{\bar \chi^0}}

\newcommand{\FX}{{F}}

\newcommand{\scalX}{{s}}

\newcommand{\FI}[1]{{F^#1}}
\newcommand{\FIbar}[1]{{\bar F^#1}}

\preprint{TUW-16-10}

\title{Constrained superfields from an anti-D3-brane in KKLT}

\author[a]{Bert Vercnocke,}
\author[b]{and Timm Wrase}

\affiliation[a]{Institute for Theoretical Physics, University of Amsterdam,\\
Science Park 904, Postbus 94485, 1090 GL Amsterdam, The Netherlands}
\affiliation[b]{Institute for Theoretical Physics, TU Wien,\\
Wiedner Hauptstrasse 8-10/136, A-1040 Vienna, Austria}

\emailAdd{bert.vercnocke@uva.nl}
\emailAdd{timm.wrase@tuwien.ac.at}

\notoc

\abstract{The KKLT construction of dS vacua \cite{Kachru:2003aw} relies on an uplift term that arises from an anti-D3-brane. It was argued by Kachru, Pearson and Verlinde \cite{Kachru:2002gs} that this anti-D3-brane is an excited state in a supersymmetric theory since it can decay to a supersymmetric ground state. Hence the anti-D3-brane breaks supersymmetry spontaneously and one should be able to package all the world-volume fields on the anti-D3-brane into a four dimensional $\cN=1$ supersymmetric action. Here we extend previous results and identify the constrained superfields that correspond to all the degrees of freedom on the anti-D3-brane. In particular, we show explicitly that the four 4D worldvolume spinors give rise to constrained chiral multiplets $S$ and $Y^i$, $i=1,2,3$ that satisfy $S^2=SY^i=0$. We also conjecture (and provide evidence in a forthcoming publication) that the vector field $A_\mu$ and the three scalars $\phi^i$ give rise to a field strength multiplet $W_\alpha$ and three chiral multiplets $H^i$ that satisfy the constraints $S W_\alpha= \bar{D}_{\dot \alpha} (S \bar H^i)=0$. This is the first time that such constrained multiplets appear in string theory constructions.
}

\begin{document}

\maketitle

\section{Introduction}
The first construction of dS vacua in string theory were obtained in 2003 by Kachru, Kallosh, Linde and Trivedi (KKLT) \cite{Kachru:2003aw}. They stabilized all closed string moduli in a supersymmetric AdS vacuum and then added one or a stack of anti-D3-branes that are localized at the bottom of a warped throat. The extra contribution from these anti-D3-branes can uplift the minimum of the scalar potential to a positive value, i.e.\ a dS vacuum. The anti-D3-branes are therefore clearly breaking supersymmetry and there has been sometimes a confusion about whether this breaking is explicit or spontaneous. The reason for this is that the anti-D3-brane uplift term was not expressed in terms of the four dimensional $\cN=1$ {\K} and superpotential. However, already in 2001 it was shown by Kachru, Pearson and Verlinde \cite{Kachru:2002gs} that anti-D3-branes in the corresponding flux background can decay to a supersymmetric ground state. Also recent holographic studies indicate that the anti-D3-branes break supersymmetry spontaneously \cite{Bertolini:2015hua}. Hence it should be possible to package the uplift term into the four dimensional {\K} and superpotential and/or a D-term. 

This rewriting of the uplift term was accomplished only very recently in \cite{Ferrara:2014kva, Kallosh:2014wsa, Bergshoeff:2015jxa} (see also \cite{McGuirk:2012sb} for earlier work in this direction). The crucial ingredient that was used in these papers is a nilpotent chiral superfield, which is a regular chiral superfield  $S = s+\sqrt{2}\theta \psi + \theta^2 F_s$ that satisfies the constraint \footnote{We will use Majorana spinors unless stated otherwise and the notation from the book \emph{Supergravity} by Freedman and Van Proeyen \cite{freedman2012supergravity}. We will also use the shorthand notation $\psi^2 = \bar \psi P_L \psi$, $\bar \psi^2 = \bar \psi P_R \psi$ and $\bar \theta P_L \psi = \theta\psi$.}
\be
S^2=(s+\sqrt{2}\theta \psi + \theta^2 F_s)^2=s^2+2\sqrt{2} s \theta \psi +\theta^2(2s F_s - \psi^2)=0\,.
\ee
The three resulting constraints above are all solved by
\be\label{eq:s}
s=\frac{\psi^2}{2F_s}\,.
\ee 
This means that the nilpotent chiral multiplet contains no scalar which then requires that supersymmetry is non-linearly realized. A corresponding supersymmetric action for a single fermion was first written down by Volkov and Akulov (VA) \cite{Volkov:1972jx, Volkov:1973ix}, while the connection between the fermion of VA and the one in the nilpotent chiral superfields was established in \cite{Rocek:1978nb, Ivanov:1978mx, Lindstrom:1979kq, Casalbuoni:1988xh, Komargodski:2009rz}. A variety of different supersymmetric actions for a single fermion were written down in the past but it was shown in \cite{Kuzenko:2010ef, Kuzenko:2011tj} that all of these are related by non-linear field redefinitions and that there is a unique action, that is given by
\be\label{eq:SVA}
S =- \int E^0 \w E^1 \w E^2 \w E^3 \qquad \text{with} \qquad E^\mu = dx^\mu + \bar \lambda^0 \gamma^\mu d \lambda^0\,.
\ee
Here $\lambda^0$ is the single fermion and the action is invariant under the following non-linear supersymmetry transformation
\be\label{eq:VAtrafo}
\delta_{\epsilon^0} \lambda^0 = \epsilon^0 + (\bar \lambda^0 \gamma^\mu \epsilon^0) \partial_\mu \lambda^0\,.
\ee
The four dimensional $\cN=1$ supersymmetry is in this case spontaneously broken, as can be seen from the above transformation \eqref{eq:VAtrafo}, that is incompatible with a vanishing fermion, or from the constraint given in \eqref{eq:s} that requires $F_s\neq0$. The fermion $\lambda^0$ is the massless Goldstino we expect in a supersymmetric theory with broken supersymmetry.

Using the nilpotent chiral superfield $S$, the authors of \cite{Ferrara:2014kva} showed that the anti-D3-brane uplift term can be packaged into the {\K} and superpotential of the four dimensional $\cN=1$ theory. They also started to clarify the connection between the nilpotent chiral superfield and the action of the anti-D3-brane. This relation was then made explicit in \cite{Kallosh:2014wsa, Bergshoeff:2015jxa} as follows: To simplify the task the authors of \cite{Kallosh:2014wsa, Bergshoeff:2015jxa} studied a single anti-D3-brane on top of an O3-plane. This removes all the bosonic worldvolume degrees of freedom and leaves only a sixteen component 10d Majorana-Weyl (MW) spinor $\lambda$ on the anti-D3-brane. This 16 component MW spinor can be decomposed into four 4D Majorana spinors $\lambda^\alpha$, $\alpha=0,1,2,3$, where $\lambda^0$ is a singlet and $\lambda^i$, for $i=1,2,3$, a triplet under the $SU(3)$ group acting on the three complex transverse directions of the anti-D3-brane. The action for this anti-D3-brane on top of the O3-plane in flat space was found to be \cite{Kallosh:2014wsa}
\be\label{eq:SD3O3flat}
S^{\overline{\rm D3}} =-2 \int \tilde{E}^0 \w \tilde{E}^1 \w \tilde{E}^2 \w \tilde{E}^3 \qquad \text{with} \qquad \tilde{E}^\mu = dx^\mu + \sum_{\alpha=0}^3 \bar \lambda^\alpha \gamma^\mu d \lambda^\alpha\,.
\ee
In flat space with an O3-orientifold the anti-D3-brane breaks the four dimensional $\cN=4$ supersymmetry spontaneously and the four $\lambda^\alpha$ correspond to the four Goldstinos. However, upon placing the anti-D3-brane into the flux background of Giddings, Kachru and Polchinski (GKP) \cite{Giddings:2001yu} that preserves only four dimensional $\cN=1$ supersymmetry, one can show that the $\lambda^i$ get a mass and decouple from the low energy effective action \cite{Bergshoeff:2015jxa}. The anti-D3-brane action \eqref{eq:SD3O3flat} then reduces to the Volkov-Akulov action given in \eqref{eq:SVA}. This shows explicitly that the low energy degrees of freedom on an anti-D3-brane on top of an O3-plane in the GKP flux background are given by a nilpotent chiral superfield $S$.

In this paper we will show how one can also package the $\lambda^i$ into constrained chiral multiplets $Y^i$ that satisfy $S Y^i=0$. In section \ref{sec:D3O3flat}, we will review the results for an anti-D3-brane on top of an O3-plane in flat space and discuss its non-linear supersymmetries. Then we review the extension to a supersymmetric GKP flux background in section \ref{sec:D3O3GKP}. The identification with a 4D supersymmetric action with constrained superfields $S$ and $Y^i$ is made explicitly in section \ref{sec:D3O34d}. We conclude and provide an outlook in section \ref{sec:conclusions} where we conjecture how to package the bosonic worldvolume fields on an anti-D3-brane into constrained 4D $\cN=1$ multiplets. This conjecture will be proven in the forthcoming paper \cite{paper2}. Appendix \ref{app:conventions} lists our spinor conventions and appendix \ref{app:nonlinear} provides several important technical details.

\section{The anti-D3-brane in flat space}\label{sec:D3O3flat}
The action for a single D-brane or a stack of D-branes including all fermionic terms is rather complicated and in many cases not even known. In particular, the action for a stack of D-branes in flat space is not known to all orders in the worldvolume fermions and the action for a single D-brane in a flux background is only known to quadratic order in the fermions. Nevertheless, one can already get a lot of insight out of these leading terms and the supersymmetry transformations of the worldvolume fields. Also for the simplest example of a single D-brane in flat space the action is known to all orders in the fermions which allows one to perform checks beyond the leading order. For such a D-brane in flat space 16 of the spacetime supersymmetries are realized linearly while the other ones are non-linearly realized and spontaneously broken (see for example page 140 of the textbook \cite{Polchinski:1998rr}). 

We are particularly interested in the non-linearly realized supersymmetries and their supersymmetry breaking. In a compactification of type IIB string theory with a standard O3/O7-orientifold projection the linearly realized supersymmetries of an anti-D3-brane are projected out by the orientifold projection and only the non-linearly realized and spontaneously broken supersymmetries remain. So these are clearly the relevant ones for the KKLT construction. To simplify the analysis of the worldvolume fields, it was suggested in \cite{Kallosh:2014wsa} to place a single anti-D3-brane on top of an O3-plane in flat space. This setup was analyzed in \cite{Sugimoto:1999tx, Antoniadis:1999xk, Uranga:1999ib} were it was shown that it is stable and that the orientifold projection removes the worldvolume bosons, i.e. the vector $A_\mu$ and the scalars $\phi^i$, from the worldvolume theory of the anti-D3-brane. We now review the detailed purely fermionic action of the anti-D3-brane in this setup and spell out the 16 non-linearly realized supersymmetries.

The action for an anti-D3-brane in flat space, including all the fermionic terms, is given by\footnote{We are using the results and notation of \cite{Bergshoeff:2013pia}. The D-brane action and its transformations are given in appendix A of that paper. We set $\alpha'=1$.}
\be\label{eq:action}
S^{\overline{\rm D3}} = S_{\text{DBI}} + S_{\text{WZ}} = - \int d^4 x \sqrt{-\text{det}(G_{\mu\nu} + \cF _{\mu\nu})} -  \int \Omega_4\,.
\ee
We denote the longitudinal and transverse coordinates as 
\be
X^M = \{ X^m, \phi_r^I \}\,,\qquad M=0,1,\ldots, 9\,, \qquad m=0,1,2,3,\qquad I=1,2,3,4,5,6\,,
\ee
where $m$ refers to the worldvolume coordinates and $I$ to the six real transverse coordinates, which we can write as three complex directions $\phi^i=\phi_r^{2i-1}+\rmi \phi_r^{2i}$, $i=1,2,3$. The $\phi^i$ are the scalar fields that control the position of the anti-D3-brane. The metric including fermionic terms is given by
\be
G_{\mu\nu} = \eta_{mn} \Pi^m_\mu \Pi^n_\nu + \delta_{IJ} \Pi^I_\mu \Pi^J_\nu\,, \qquad \Pi^m_\mu = \partial_\mu X^m-\bar \theta \Gamma^m \partial_\mu \theta\,, \qquad \Pi^I_\mu = \partial_\mu \phi_r^I - \bar \theta \Gamma^I \partial_\mu \theta\,,
\ee
where $\eta_{mn}$ is the Minkowski metric, $\Gamma^M$ are 10D gamma matrices and $\theta^\varsigma$, $\varsigma=1,2$, denotes a doublet of 16 components MW spinors of the same chirality so that $\bar\theta_\varsigma = \{\theta_1^T C, \theta_2^T C\}$ with $C$ being the charge conjugation matrix. The index $\varsigma$ will be contracted with the identity matrix or Pauli matrices $\sigma_a$, $a=1,2,3$. When it is clear from the context, we will omit this index as well as the identity matrix. We will always omit the spinorial indices.

The Born-Infeld field strength $\cF_{\mu\nu}$ is given by
\be
\cF_{\mu\nu} = F_{\mu\nu} -b_{\mu\nu}\,, \qquad b_{\mu\nu}= \bar \theta \sigma_3 \Gamma_M \partial_\mu \theta \lp \partial_\nu X^M-\frac12 \bar \theta \Gamma^M \partial_\nu \theta \rp - (\mu \leftrightarrow \nu)\,,
\ee
where $F_{\mu\nu} = \partial_\mu A_\nu -\partial_\nu A_\mu$ is the field strength of the worldvolume gauge field $A_\mu$. Lastly, the 4-form $\Omega_4$ is defined via a closed 5-form
\be \label{eq:I5}
I_5 = d \Omega_4 = d\bar \theta \lp \sigma_1 \cF \hat{\Gamma} + \rmi \sigma_2 \frac{\hat{\Gamma}^3}{3!} \rp d \theta\,,\qquad \hat{\Gamma} = \Gamma_M \Pi^M = \Gamma_M(dX^M+\bar\theta \Gamma^M d\theta)\,,
\ee
where wedge products are implicit and the plus sign in the last equation above is explained on page 5 of \cite{Aganagic:1996nn}.

With this information at hand one can explicitly spell out the component action for an anti-D3-brane but this is rather cumbersome, so the authors of \cite{Kallosh:2014wsa} placed the anti-D3-brane on top of an O3-plane that extends along the first four spacetime directions. The orientifold projection removes the vector field $A_\mu$ and the scalars $\phi^i$. Furthermore, it constraints the fermion doubled to satisfy
\be\label{eq:thetaconstraint}
(1+\rmi \sigma_2 \Gamma_{0123}) \theta = 0 \qquad \Leftrightarrow \qquad \theta^1 = -\Gamma_{0123} \theta^2\,.
\ee
After this orientifold truncation the $\kappa$-symmetry disappears (see \cite{Bergshoeff:1999bx}) and we are left with the 16 component MW spinor $\lambda =\sqrt{2} \theta^1 =-\sqrt{2} \Gamma_{0123} \theta^2$. The DBI and WZ-term are then equal and the anti-D3-brane action is given by \cite{Kallosh:2014wsa}
\be
S^{\overline{\text{D3}}} = -2 \int \tilde{E}^0 \w \tilde{E}^1\w \tilde{E}^2\w \tilde{E}^3\,, \qquad \tilde{E}^m = dX^m + \bar \lambda \Gamma^m d \lambda\,.
\ee
This action is invariant under 16 non-linearly realized supersymmetries and the corresponding transformations are
\be
\delta_\epsilon \lambda= \epsilon\,, \qquad \delta_\epsilon X^m =- \bar \lambda \Gamma^m \epsilon\,, 
\ee
where $\epsilon$ is a 16 component MW spinor. After fixing the diffeomorphism symmetry so that $X^m(x)= \delta^m_\mu x^\mu$, we are left with a non-linear supersymmetry transformation that is very similar to the above in equation \eqref{eq:VAtrafo}, namely
\be\label{eq:VAtrafoN4}
\delta_\epsilon \lambda = \epsilon + (\bar \lambda \Gamma^\mu \epsilon) \partial_\mu \lambda\,.
\ee
We can rewrite the above action in terms of four 4D spinors $\lambda^\alpha$ and 4D gamma matrices $\gamma^\mu$ which leads to equation \eqref{eq:SD3O3flat}
\be\label{eq:SD3O3flat2}
S^{\overline{\text{D3}}} = -2 \int \tilde{E}^0 \w \tilde{E}^1 \w \tilde{E}^2 \w \tilde{E}^3 \qquad \text{with} \qquad \tilde{E}^\mu = dx^\mu + \sum_{\alpha=0}^3 \bar \lambda^\alpha \gamma^\mu d \lambda^\alpha\,.
\ee
The transformation rules for the four spinors $\lambda^\alpha$ are now
\be\label{eq:trafoSU4}
\delta_{\epsilon} \lambda^\alpha = \delta^\alpha_\beta \epsilon^\beta + \sum_\beta (\bar \lambda^\beta \gamma^\mu \epsilon^\beta) \partial_\mu \lambda^\alpha\,.
\ee
In the non-trivial backgrounds of GKP \cite{Giddings:2001yu} or KKLT \cite{Kachru:2003aw} the supersymmetries $\epsilon^i$, $i=1,2,3$ are broken by the background so we will be particularly interested in the 4D $\cN=1$ supersymmetry generated by $\epsilon^0$. Under this symmetry we have the following transformation rules
\ba\label{eq:trafoferm}
\delta_{\epsilon^0} \lambda^0 &=& \epsilon^0 + (\bar \lambda^0 \gamma^\mu \epsilon^0) \partial_\mu \lambda^0\,,\cr
\delta_{\epsilon^0} \lambda^i &=& (\bar \lambda^0 \gamma^\mu \epsilon^0) \partial_\mu \lambda^i\,.
\ea
Since it will become important for us later, let us discuss how the above transformations relate different terms in the action \eqref{eq:SD3O3flat2}. Expanding this action to quadratic order in fermions we get
\be
S^{\overline{\text{D3}}} = -2\int \lp 1 + \bar \lambda^0 \gamma^\mu \partial_\mu \lambda^0+\sum_{i=1}^3 \bar \lambda^i \gamma^\mu \partial_\mu \lambda^i + \ldots \rp\,,
\ee
where at higher order we can have terms that only contain $\lambda^0$, only contain the $\lambda^i$ or a mixture of both. The index structure of these terms is not important here, so let us denotes these terms schematically by $(\bar \lambda^0 \gamma \partial \lambda^0)^{p_1} (\bar \lambda^i \gamma \partial \lambda^i)^{p_2}$, where $p_1,p_2=0,1,2,3,4$, $p_1+p_2\leq 4$. Up to total derivatives, the transformations given in \eqref{eq:trafoferm} relate terms that only involve $\lambda^0$ as follows
\be
\ldots \rightarrow (\bar\epsilon^0 \gamma \partial \lambda^0)(\bar \lambda^0 \gamma \partial \lambda^0)^{p_1-1} \leftarrow \boxed{(\bar \lambda^0 \gamma \partial \lambda^0)^{p_1}} \rightarrow (\bar\epsilon^0 \gamma \partial \lambda^0)(\bar \lambda^0 \gamma \partial \lambda^0)^{p_1} \leftarrow \boxed{(\bar \lambda^0 \gamma \partial \lambda^0)^{p_1+1}} \rightarrow \ldots\nn
\ee
This means that the terms $(\bar \lambda^0 \gamma \partial \lambda^0)^{p_1}$ for $p_1=1,2,3,4$ are all related by the non-linear supersymmetry transformation \eqref{eq:trafoferm}. The invariance of the action under this transformation requires all these higher order derivative terms to be present and they have to have exactly the form obtained from expanding the Volkov-Akulov action in equation \eqref{eq:SVA}.\footnote{We have explicitly checked this keeping track of the indices on the gamma matrices and partial derivatives.} 

This is different for the terms that only contain the spinors $\lambda^i$. We find the mapping
\be
\boxed{(\bar \lambda^i \gamma \partial \lambda^i)^{p_2}} \rightarrow (\bar\epsilon^0 \gamma \partial \lambda^0)(\bar \lambda^i \gamma \partial \lambda^i)^{p_2}\leftarrow \boxed{(\bar\lambda^0 \gamma \partial \lambda^0)(\bar \lambda^i \gamma \partial \lambda^i)^{p_2}}\rightarrow  (\bar\epsilon^0 \gamma \partial \lambda^0)(\bar\lambda^0 \gamma \partial \lambda^0)(\bar \lambda^i \gamma \partial \lambda^i)^{p_2}\leftarrow \ldots\nn
\ee
Thus we see that the transformation \eqref{eq:trafoferm} relates the term $(\bar \lambda^i \gamma \partial \lambda^i)^{p_2}$ for fixed $p_2$ to all the terms $(\bar \lambda^0 \gamma \partial \lambda^0)^{p_1}(\bar \lambda^i \gamma \partial \lambda^i)^{p_2}$ for an arbitrary $p_1$. So for example, starting with the standard kinetic term $\bar \lambda^i \gamma^\mu\partial_\mu \lambda^i$, the non-linear supersymmetry transformation will only constrain the terms quadratic in the $\lambda^i$ like $(\bar \lambda^0 \gamma \partial \lambda^0)^{p_1}(\bar \lambda^i \gamma \partial \lambda^i)$ for arbitrary $p_1$. However, higher derivative terms like $(\bar \lambda^i \gamma \partial \lambda^i)^2$ are not constrained by the leading order terms in a derivative expansion.

Below in subsection \ref{sec:D3O34d} we will start with the standard supersymmetric action for the four chiral multiplets $S$ and $Y^i$. This action contains only terms that are leading order in derivatives and when we impose the constraints $S^2=SY^i=0$, we will only find the higher derivative terms that are related to the standard kinetic terms by the non-linear supersymmetry transformations \eqref{eq:trafoferm}. In hindsight this might have been expected but it is somewhat different from a single nilpotent chiral multiplet $S$, for which the action, including all higher derivative terms $(\bar \lambda^0 \gamma \partial \lambda^0)^{p_1}$, is fixed by the symmetry. The explanation for this difference is that the action in equation \eqref{eq:SD3O3flat2} is actually invariant under the non-linearly realized $\cN=4$ supersymmetry given in equation \eqref{eq:trafoSU4} and not just the non-linearly realized $\cN=1$ supersymmetry in \eqref{eq:trafoferm}. If we were to impose this larger symmetry group, then we would of course reproduce the full anti-D3-brane action starting from the action for $S$ and $Y^i$.

In the next section we review the anti-D3-brane action in the non-trivial flux background of GKP \cite{Giddings:2001yu}. The GKP  background breaks the 12 supercharges $\epsilon^i$, $i=1,2,3$ and the fluxes give a mass to the three spinors $\lambda^i$.

\section{The anti-D3-brane in a GKP flux background}\label{sec:D3O3GKP}
In \cite{Bergshoeff:2015jxa} (see also \cite{McGuirk:2012sb} for earlier related work) the above analysis was extend to an anti-D3-brane on top of an O3-plane in a GKP flux background \cite{Giddings:2001yu}. For an anti-D3-brane in a general background the action is only know to quadratic order in the fermions. This leading order fermionic part of the action is given by \cite{Marolf:2003ye, Marolf:2003vf, Martucci:2005rb}
\be
S_f^{\overline{\text{D3}}} = \frac{T_3}{2} \int d^4x\, e^{-\phi} \sqrt{-\text{det}\lp g+ \cF\rp}\, \bar \theta (1-\Gamma_{\overline{\text{D3}}}) \ls \lp g+ \Gamma_{10} \sigma_3 \cF\rp^{-1\ \mu\nu} \Gamma_\mu D_\nu-\Delta \rs \theta\,.
\ee
Here $T_3$ is the brane tension, $\phi$ the dilaton, $g_{\mu\nu}$ the pullback of the background metric and
\be
\Gamma_{\overline{\text{D3}}} = -\frac{\sigma_1 \Gamma_{0123}}{\sqrt{- \text{det}(g+ \cF)}}\lp 1 + \frac{\sigma_3}{2} \Gamma^{\mu_1 \mu_2} \cF_{\mu_1\mu_2} + \frac{1}{8} \Gamma^{\mu_1 \mu_2\mu_3\mu_4} \cF_{\mu_1\mu_2}\cF_{\mu_3\mu_4}   \rp\,.
\ee
$D_\mu$ and $\Delta$ depend on the background fluxes and are related to the supersymmetry transformations of two of the background fields, namely the gravitino and dilatino, respectively. They are explicitly given by
\ba
D_M &=& \nabla_M + \frac{1}{8} H_{MNP}\Gamma^{NP}\sigma_3 \cr
&&+ \frac{1}{8} e^\phi \lp  F_{N}\Gamma^{N}(\rmi \sigma_2) +\frac{1}{3!} F_{NPQ}\Gamma^{NPQ} \sigma_1 +\frac{1}{2\cdot5!} F_{NPQRT}\Gamma^{NPQRT} (\rmi \sigma_2) \rp \Gamma_M\,,\cr
\Delta &=& \frac12 \Gamma^M \partial_M \phi+\frac{1}{24} H_{MNP}\Gamma^{MNP} \sigma_3 -\frac{e^\phi}{2} \lp F_M \Gamma^M (\rmi \sigma_2) + \frac{1}{2\cdot 3!} F_{MNP}\Gamma^{MNP} \sigma_1 \rp\,, \qquad
\ea
where $H$ denotes the NSNS 3-form flux and the $F$'s the RR fluxes of the type IIB background.

Again for an anti-D3-brane on top of an O3-plane, the gauge field $A_\mu$ and the scalars $\phi^i$ are projected out from the spectrum. Specializing further to a GKP background with primitive (2,1)-flux that preserves $\cN=1$ supersymmetry in four dimensions, the authors of \cite{Bergshoeff:2015jxa} find that the action reduces to
\be\label{eq:actionGKP}
S_f^{\overline{\text{D3}}} = 2 T_3 \int d^4x\, e^{4A_0-\phi} \ls \bar \lambda_-^0 \gamma^\mu \nabla_\mu \lambda_+^0 + \delta_{i\jb} \bar \lambda_-^{\jb} \gamma^\mu \nabla_\mu \lambda_+^i + \frac12 m_{ij} \bar \lambda_+^i \lambda_+^j + \frac12 \bar m_{\ib\jb} \bar \lambda_-^\ib \lambda_-^\jb \rs\,,
\ee
where the subscripts $\pm$ denote 4D Weyl spinors that satisfy $\lambda_\pm = \frac12 (1 \pm \rmi \gamma_{0123}) \lambda$ and $e^{A_0}$ is the warp factor evaluated at the anti-D3-brane location. The mass matrix for the $\lambda^i$ is given by
\be\label{eq:massmatrix}
m_{ij} = \frac{\sqrt{2}}{8} \rmi e^\phi (e^w_i e^t_j+e^w_j e^t_i) \Omega_{uvw} g^{u\bar u} g^{v \bar v} \bar G^{\text{ISD}}_{t \bar u \bar v}\,, \qquad u,v,w,t=1,2,3\,.
\ee
The matrix $\bar{m}_{\ib\jb}$ is the complex conjugate of $m_{ij}$, $\Omega$ is the holomorphic 3-form, $g_{u\bar v}$ the CY$_3$-metric, $e_i^u$ the corresponding vielbein and the complexified 3-form flux is defined as
\be
\bar G_3 = F_3 + \rmi e^{-\phi} H_3\,,\qquad \bar G_3^{\text{ISD}} = \frac12 \lp \bar G_3 + \rmi *_6 \bar G_3 \rp\,.
\ee
Thus in this setting where the background only preserves the $\cN=1$ supersymmetry corresponding to $\epsilon^0$, one finds that the $SU(3)$ triplet $\lambda^i$ receives a mass, while the Goldstino $\lambda^0$ remains of course massless.\footnote{The authors of \cite{Bergshoeff:2015jxa} also showed that $(3,0)$ flux gives a mass to $\lambda^0$. Then the background breaks all the supersymmetry and $\lambda^0$ is no longer the Goldstino. In the KKLT setup there is one additional contribution which comes from Euclidean D3-branes or gaugino condensation on a stack of D7-branes. Since both of these extra sources are localized in the internal dimension, we expect that they generically do not affect our anti-D3-brane action. However, these new ingredients lead to supersymmetric AdS vacua even in the presence of (3,0)-flux. Thus the anti-D3-brane is solely responsible for the spontaneous breaking of the 4D $\cN=1$ supersymmetry in the KKLT setup and the SU(3) singlet worldvolume fermion $\lambda^0$ is the Goldstino.}

Since the action in a non-trivial background is only known to leading quadratic order in the fermions, the supersymmetry transformations are likewise not known beyond the leading order and are given by \cite{Marolf:2003vf}
\ba
\delta_{\epsilon^0} \lambda^0 &=& \epsilon^0 + \mathcal{O}\lp(\lambda^\alpha)^2\rp\,,\cr
\delta_{\epsilon^0} \lambda^i &=&\mathcal{O}\lp(\lambda^\alpha)^2\rp\,.
\ea
These transformations are of course not very restrictive and with the leading order action and transformations for the anti-D3-brane we can in principle only achieve a leading order matching with a 4D SUSY action. However, in the limit of vanishing background fluxes the above anti-D3-brane action reduces to the flat space action and we can think of it as a deformation of the flat space result. This deformation corresponds to turning on the (2,1) ISD-flux which gives rise to the mass matrix for the fermions $\lambda^i$. In the next subsection we spell out the SUSY action that reproduces the anti-D3-brane action in flat space as well as the deformation that corresponds to the mass matrix $m_{ij}$.

\section{The four dimensional \texorpdfstring{$\cN=1$}{} action}\label{sec:D3O34d}
As discussed above, an anti-D3-brane on top of an O3 orientifold plane in a supersymmetric GKP background has one massless 4D fermion $\lambda^0$ and three massive fermions $\lambda^i$, $i=1,2,3$ with a non-linearly realized supersymmetry (see equation \eqref{eq:trafoferm})
\ba
\delta_{\epsilon^0} \lambda^0 &=& \epsilon^0 + (\bar \lambda^0 \gamma^\mu \epsilon^0) \partial_\mu \lambda^0\,,\cr
\delta_{\epsilon^0} \lambda^i &=& (\bar \lambda^0 \gamma^\mu \epsilon^0) \partial_\mu \lambda^i\,.\label{eq:trafoferm2}
\ea
These are the transformations of a Goldstino $\lambda^{0}$ and a triplet of fermions $\lambda^i$ that transform in the standard way under a non-linearly realized $\cN=1$ supersymmetry \cite{Ivanov:1977my,Ivanov:1978mx}.

We now discuss how to fit these fermions in the standard description of $\cN=1$ supersymmetry. For that it is important to realize that any superfield can be turned into a non-linearly transforming one. For a review of this procedure and references, see appendix \ref{app:nonlinear}. So the question is not whether we can fit the spinors into non-linearly transforming multiplets, but rather in which multiplets they will sit. 

For this discussion, we use the language of constrained superfields. The constraints act as a way to eliminate components of superfields. We find that for each of the four fermions we need one chiral superfield, whose bosonic degrees of freedom are removed by a constraint.
The Goldstino $\lambda^0$ is described by a nilpotent chiral superfield $S$, obeying $S^2 =0$. We now show that the $SU(3)$ triplet $\lambda^i$ is described by a triplet of chiral superfields $Y^i = 0$, such that 
\begin{equation}
 S Y^i = 0\,.
\end{equation}
For that purpose it is more convenient to write the four-dimensional Majorana spinors $\lambda^0,\lambda^i$ in terms of complex two-spinors \footnote{We are following the conventions of Wess and Bagger \cite{Wess:1992cp} for the rest of this paper.}
\begin{equation}
\lambda^0 = \begin{pmatrix}\fermXb\\\fermXbbar\end{pmatrix}\,,\qquad \lambda^i = \begin{pmatrix}\fermb{i}\\\bar \chi^i\end{pmatrix}\,.
\end{equation}
We write the chiral superfields $S,Y^i$ in components 
\begin{equation}
 S = \scalX + \sqrt 2 \theta \fermX + \theta^2 \FX_s\,,\qquad Y^i = y^i + \sqrt 2 \theta \ferm{i} + \theta^2\FI{i}\,,
\end{equation}
and the constraints $S^2=S Y^i = 0$ give the relations
\ba
\scalX &=& \frac{(\fermX)^2}{2\FX_s}\,,\nonumber\\
y^i &=&\frac {\psi^0\ferm{i}}{\FX_s} - \frac{(\psi^0)^2}{2\FX_s^2} \FI{i}\,.\label{eq:XandXYconstraints}
\ea
We stress that the spinor components $\psi^0$ and $\psi^i$ of $S$ and $Y^i$ are not directly those of the original fields $\lambda^0$ and $\lambda^i$ (or $\chi^0$ and $\chi^i$). One has to perform a field redefinition involving higher order terms of the fermion singlet and triplet to obtain combinations of fields that transform properly under the non-linearly realized supersymmetry, i.e. like the $\lambda^0$ and $\lambda^i$ in equation \eqref{eq:trafoferm2}. We explain this in detail in appendix \ref{app:nonlinear}, and here we just give the redefinitions that lead to the fermions with the correct non-linear transformations:
\ba
\fermXb(x) &=& \frac{\fermX(\hat x)}{\sqrt 2 \FX_s(\hat x)}\,,\label{eq:field_redef_Yi}\\
\fermb{i}(x) &=& \ferm{i}(\hat x) - \sqrt 2 \FI{i}(\hat x) \fermXb (x) - \sqrt 2 \rmi (\sigma^\mu \fermXbbar(x)) \left[\sqrt2  \ferm{i}(\hat x)\partial_\mu \fermXb(x) -  2\FI{i}(\hat x)\fermXb (x)\partial_\mu \fermXb(x) \right]\,,\nonumber
\ea
with the field redefinition implicitly defined through
\be
\hat x^\mu =  x^\mu + \rmi \fermXb (x) \sigma^\mu  \fermXbbar(x)\,.
\ee
For a two-dimensional spinor SUSY parameter $\epsilon^0$ defined as $\epsilon^0 = (\xi,\bar \xi)^T$, the SUSY transformation of the 2-component spinors is
\ba\label{eq:trafoferm3}
\delta_\xi \fermXb &=&  \xi- \rmi  (\fermXb \sigma^\mu \bar \xi - \xi \sigma^\mu  \fermXbbar) \partial_\mu \fermXb\,, \cr
\delta_\xi \fermb{i}   &=&  -\rmi (\fermXb \sigma^\mu \bar \xi - \xi \sigma^\mu  \fermXbbar) \partial_\mu \fermb{i}\,,
\ea
which is the same as \eqref{eq:trafoferm2}. Thus we have identified the implicit field redefinitions given in equation \eqref{eq:field_redef_Yi} that map the spinors $\psi^0$ and $\psi^i$ in $S$ and $Y^i$ to the spinors $\lambda^0$ and $\lambda^i$ on the anti-D3-brane. 

Let us now identify the 4D $\cN=1$ supersymmetric action for the $S$ and $Y^i$ superfields that reproduces the anti-D3-brane action. The coupling of $S$ and a single constrained field $Y$ to supergravity was recently studied in detail in \cite{Dall'Agata:2015zla}, which substantially simplifies our task.

We take the following K\"ahler and superpotential
\begin{equation}
K =c  S \bar S + \delta_{i\ib} Y^i \bar Y^{\ib}\,, \qquad W = f S + g_i Y^i  + h_{ij}  Y^iY^j\,,
\end{equation}
where $c \in \mathbb{R}$ and $f, g_i, h_{ij} \in \mathbb{C}$. The Lagrangian up to total derivatives is
\ba
{\cal L}  &=& \int d^2 \theta d^2\bar \theta K + \int d^2 \theta  W + \int d^2 \bar \theta \bar W\nonumber\\
&=& \rmi c\, \partial_m \fermXbar  \bar \sigma^m \fermX  + c\, \bar \scalX \partial^2 \scalX+\rmi  \partial_m   \fermbar{\ib} \bar\sigma^m \ferm{i}\delta_{i\ib} +  \bar y^{\ib} \partial^2 y^i\delta_{i\ib}
\nonumber\\
&&+\bar \FX_s \FX_s + \FIbar{\ib} \FI{i}\delta_{i\ib}+[f \FX_s + (g_i +2h_{ij}y^j) \FI{i}- h_{ij}\ferm{i}\ferm{j}+ {\rm h.c.}]\,.
\ea
We now use the constraints $S^2 = 0$, $S Y^i = 0$ to make the scalars dependent variables and replace them with the fermion bilinears given in equation \eqref{eq:XandXYconstraints} to get
\ba\label{eq:globalN=1chiralaction}
 {\cal L} &=& \rmi c\, \partial_m \fermXbar  \bar \sigma^m \fermX  +c\, \frac{(\fermXbar)^2}{2\bar F_s} \partial^2 \lp\frac{(\fermX)^2}{2\FX_s} \rp+\rmi\partial_m   \fermbar{\ib} \bar\sigma^m \ferm{i} \delta_{i\ib}\cr
 &&+ \lp \frac {\bar \psi^0 \bar \psi^{\ib}}{\bar F_s} - \frac{(\bar \psi^0)^2}{2\bar{\FX}_s^2} \bar F^{\ib} \rp \partial^2 \lp \frac {\psi^0 \ferm{i} }{\FX_s}- \frac{(\psi^0)^2}{2\FX_s^2} \FI{i}\rp \delta_{i\ib}+\bar \FX_s \FX_s + \FIbar{\ib} \FI{i}\delta_{i\ib} \cr
 &&+\ls f \FX_s+ \lp g_i +2h_{ij}\lp \frac {\psi^0 \ferm{j}}{\FX_s}- \frac{(\psi^0)^2}{2\FX_s^2} \FI{j}\rp\rp \FI{i}- h_{ij}\ferm{i}\ferm{j}+ {\rm h.c.}\rs. \qquad \quad
\ea

The F-term equations of motion are
\ba
0 &=& \FX_s + \bar f -c\, \frac{(\fermXbar)^2}{2\bar F_s^2} \partial^2 \left(\frac{(\fermX) ^2}{2 F_s}\right) + \left(-\frac{\fermXbar\fermbar{\ib} }{\bar F_s^2}+ \frac{(\fermXbar)^2 }{ \bar F_s^3}\FIbar{\ib}\right)\partial^2 \left(\frac{\fermX\ferm{i} }{F_s} - \frac{(\fermX) ^2}{2F_s^2} \FI{i}\right)\delta_{i\ib} \cr
&&-2 \bar h_{\ib\jb}\left(\frac{\fermXbar\fermbar{\jb} }{\bar F_s^2}- \frac{(\fermXbar)^2 }{ \bar F_s^3}\FIbar{\jb}\right) \bar F^{\ib}\,,\cr 
0&=& F^i\delta_{i\ib}+ \bar g_{\ib} +2 \bar h_{\ib\jb} \lp \frac {\bar \psi^0 \fermbar{\jb}}{\bar \FX_s}- \frac{(\bar\psi^0)^2}{\bar \FX_s^2} \bar F^{\jb} \rp- \frac{(\fermXbar)^2}{2\bar F_s^2}  \partial^2 \left(\frac{\fermX \ferm{i} }{F_s} - \frac{(\fermX) ^2}{2F_s^2} \FI{i}\right)\delta_{i\ib}\,.\label{eq:Fterms}
\ea
We want to match to the anti-D3-brane actions given in equation \eqref{eq:SD3O3flat2} for flat space and in equation \eqref{eq:actionGKP} for a GKP background. For both backgrounds the Goldstino is $\lambda^0$ so that we only want the SUSY breaking to arise from $F_s$ developing a vev. Thus we choose $g_i=0$. We can then solve the F-term equations iteratively in a fermion expansion. In particular, at lowest order we find
\ba
F_s &=& -\bar f + \mathcal{O}(\psi^4)\,,\cr
F^i &=& \mathcal{O}(\psi^2)\,.
\ea
Now we can likewise expand the field redefinitions in equation \eqref{eq:field_redef_Yi} to lowest order and find
\ba
\fermXb(x) &=& -\frac{\fermX(x)}{\sqrt 2 \bar f}+ \mathcal{O}(\psi^3)\,,\cr
\fermb{i}(x) &=& \ferm{i}(x)+ \mathcal{O}(\psi^3) \,.
\ea
This then leads to the following Lagrangian to leading order in fermions
\be\label{eq:Lleading}
{\cal L} =-f\bar f +  \rmi 2 cf\bar f \, \partial_m \bar \chi^0 \bar \sigma^m \chi^0  +\rmi\partial_m \bar \chi^{\ib} \bar\sigma^m \chi^i \delta_{i\ib}- h_{ij}\chi^i\chi^j - \bar h_{\ib\jb}\bar \chi^{\ib}\bar \chi^{\jb} + \mathcal{O}(\chi^4) \,.
\ee
Thus we see that we need to take $c=1/(2f\bar f)$ to canonically normalize the kinetic term for the $\chi^0$.

Let us now discuss how this Lagrangian in equation \eqref{eq:Lleading} reproduces the action for the anti-D3-brane on top of an O3-plane in flat space. For that purpose we have to set $h_{ij}=0$. We find that the field redefinitions in equation \eqref{eq:field_redef_Yi} give a match of the kinetic terms. All terms determined by the non-linearly realized 4D $\cN=1$ supersymmetry (cf. eqs. \eqref{eq:trafoferm2}, \eqref{eq:trafoferm3}) are then matched automatically as well (see the discussion at the end of section \ref{sec:D3O3flat}). We checked explicitly the quartic terms for the Lagrangian in equation \eqref{eq:Lleading} and found a match up to terms quartic in the $\chi^i$. As mentioned above, to also match those terms we would need to demand invariance under an enhanced non-linear $\cN=4$ supersymmetry and not just the $\cN=1$ supersymmetry in equation \eqref{eq:trafoferm3}. Our results extend the work of Kuzenko and Tyler \cite{Kuzenko:2010ef} for the singlet only. They matched the Volkov-Akulov action for the fermion $\lambda^0$ (cf. equation \eqref{eq:SD3O3flat2} with $\lambda^i = 0$) to the Komargodski-Seiberg action for the nilpotent field $S$ (cf. equation \eqref{eq:globalN=1chiralaction} with $\psi^i=0$), while we also matched the terms that include $\lambda^i$ and $\psi^i$.

We also see from the leading order Lagrangian in equation \eqref{eq:Lleading} that we can turn on a mass term for the SU(3) fermion triplet. This corresponds to the anti-D3-brane in a supersymmetric GKP background that we discussed in section \ref{sec:D3O3GKP}. The matching of the Lagrangian in \eqref{eq:Lleading} and the action in \eqref{eq:actionGKP} is trivial and we see that the fermionic mass matrix $h_{ij}$ gets identified with the (2,1) ISD flux via equation \eqref{eq:massmatrix}.

\section{Conclusions and Outlook}\label{sec:conclusions}
When the anti-D3-brane sits on top of an O3-plane the worldvolume bosons, i.e.\ the gauge field $A_\mu$ and the three complex scalars $\phi^i$, are projected out. For a flat background as well as a GKP or KKLT flux background, we have shown in this paper that the remaining worldvolume spinors can be packaged into four constrained chiral $\cN=1$ supermultiplets. These constrained chiral multiplets $S$ and $Y^i$, $i=1,2,3$ satisfy the constraints $S^2=S Y^i=0$. In section \ref{sec:D3O34d}, we have spelled out the explicit 4D $\cN=1$ SUSY action in terms of the {\K} and superpotential that matches the anti-D3-brane action.

 It was shown in \cite{Kallosh:2015nia, Garcia-Etxebarria:2015lif, Retolaza:2016alb}  that such an anti-D3-brane on top of an O3-plane can arise at the bottom of warped throats (including the Klebanov-Strassler (KS) throat \cite{Klebanov:2000hb}). Thus there seems no remaining obstruction and the above findings should apply to any KKLT construction that has a warped throat like the KS throat or any other throat that allows for an O3-plane at the bottom of the throat. Likewise it was argued in \cite{Kallosh:2015nia} that one obtains the same low energy action given in equation \eqref{eq:SVA}, if one places an anti-D3-brane on top of an O7-plane, a situation that can equally arise in warped throats. While this is very satisfying, neither of these setups are generic and therefore it is useful to extend the above analysis to an anti-D3-brane is not sitting on top of an O-plane and thus none of its worldvolume fields are projected out. This involves additional worldvolume fields, namely one vector field $A_\mu$ and three complex scalars $\phi^i$. We conjecture that the vector field can be package into a constrained field strength multiplet $W_\alpha$ that satisfies $S W_\alpha =0$ and the scalars $\phi^i$ give rise to three constrained chiral multiplets $H^i$ that satisfy $\bar{D}_{\dot \alpha} (S \bar H^{\ib}) =0$. 
 
We give strong evidence for this conjecture in another publication \cite{paper2}. For now we suffice with some intuition coming from DBI actions in flat space with less supersymmetry. For the spontaneous breaking of $\cN=1$, we know that we can describe the Goldstino in terms of a nilpotent superfield $S$ satisfying $S^2=0$. Similarly, for the DBI action with $\N=2$ supersymmetry broken to $\N=1$, the $\N=2$ vector multiplet $\cal W$ is nilpotent in $\N=2$ superspace ${\cal W}^2 =0$. The $\N=2$ vector multiplet can be decomposed in $\N=1$ chiral superfields $S$ and $W_\alpha$, and the vector superfield $W_\alpha$ is the goldstone multiplet of this partial breaking. The constraint ${\cal W}^2 = 0$ gives $S = S \bar D^2 \bar S + \frac 12 W_\alpha W^\alpha$ \cite{Bagger:1996wp, Rocek:1997hi} which implies $S^2 = SW_\alpha =0$ (but the converse is not true). The DBI action in flat space we discussed has, from the four-dimensional point of view, an $\N=8$ symmetry that is spontaneously broken to $\N=4$, so we can similarly interpret the DBI action invariant under $\N=4$ linearly realized symmetries in terms of $\N=1$ superfields to derive the constraints, and possible corrections. This has only been achieved partially in the literature \cite{Tseytlin:1999dj, Bellucci:2000ft} and to the best of our knowledge there is no interpretation for the constraints on $\N=1$ superfields. The constraints $\bar{D}_{\dot \alpha} (S \bar H^{\ib}) =0$ and $SY^i=0$ are the most natural $SU(3)$ invariant constraints that give the correct non-linear transformations of the scalar and fermion triplet and as we have shown here $SY^i=0$ is indeed reproducing the anti-D3-brane action for an anti-D3-brane on top of an O3-plane.

Constrained $\cN=1$ superfields have received a lot of attention recently. They have been studied in the context of string theory in \cite{Aparicio:2015psl, Bandos:2015xnf, Dasgupta:2016prs} and their coupling to supergravity and their role in cosmology has been investigated in dozens of papers in the last two years (see the review articles \cite{Ferrara:2015exa, Ferrara:2015cwa, Schillo:2015ssx} for a partial list of references). We hope that our identification of a clear string theory origin for superfields that satisfy the constraints $S^2=S Y^i = S W_\alpha = \bar{D}_{\dot \alpha} (S \bar H^{\ib}) =0$, will provide useful in the future.

\acknowledgments
We are grateful to S. Kachru, R. Kallosh and A. Van Proeyen for enlightening discussions and collaboration on related projects. The work of TW is supported by COST MP1210. TW thanks the Department of Physics of Stanford University for the hospitality during a visit in which this work was initiated. BV acknowledges support from a Starting Grant  of the European Research Council (ERC  STG  Grant  279617).

\appendix{

\section{Spinor conventions}\label{app:conventions}

We use the conventions of \cite{freedman2012supergravity} everywhere, except in section \ref{sec:D3O34d} and appendix \ref{app:nonlinear}, where we stick to the two-component spinor conventions of Wess and Bagger \cite{Wess:1992cp}. 

We write four-dimensional Majorana spinors as
\be
\lambda = \begin{pmatrix}\chi_\alpha\\ \bar \chi^{\dot \alpha}\end{pmatrix}\,.
\ee
For Majorana spinors the conjugate is equal to $\bar \lambda = \lambda^T {\cal C}$, with $\cal C$ the charge conjugation matrix. 

The convenient choice to go to two-component spinor notation for the gamma matrices is
\begin{equation}
\gamma^0 = -\rmi \begin{pmatrix} 0& -\mathbbm{1}_2\\ -\mathbbm{1}_2&0\end{pmatrix}\,,\qquad \gamma^a= -\rmi \begin{pmatrix} 0& \sigma^a\\ -\sigma^a&0\end{pmatrix}\,,
\end{equation}
with $\sigma^a$ the three Pauli matrices. We take the charge conjugation matrix in this basis to be 
\begin{equation}
{\cal C} =\begin{pmatrix}-\varepsilon&0\\0&\varepsilon\end{pmatrix}
\end{equation}
with $\varepsilon$ the totally antisymmetric symbol normalized as $1= \varepsilon^{12} = -\varepsilon_{12}$.

For instance for the kinetic term of a Majorana spinor, we then find
\begin{equation}
 \bar \lambda \gamma^\mu \partial_\mu \lambda = -\rmi(\chi \sigma^\mu \partial_\mu \bar \chi + \bar \chi \bar \sigma^\mu \partial_\mu \chi)\,,
\end{equation}
with $\bar \sigma^a = -\sigma^a$ and $\sigma^0 = \bar \sigma^0 = -\mathbbm{1}_2$, and similarly for $\epsilon= (\xi, \bar \xi)^T$ we have 
\be
\bar \lambda \gamma^\mu \epsilon = -\rmi(\chi \sigma^\mu \bar\xi + \bar \chi \bar \sigma^\mu \xi) = -\rmi(\chi \sigma^\mu \bar\xi - \xi \sigma^\mu \bar \chi)\,.
\ee

\section{Non-linear superfields and constraints}\label{app:nonlinear}

We review how constrained superfields can describe non-linear realizations of supersymmetry. First we review how any superfield can be endowed with non-linear transformations on all components, then we give an explicit example with a chiral multiplet. Finally we review how non-linear constraints can eliminate unwanted components of the non-linearly transforming superfield, such that one can effectively have superpartner-less fields.

\subsection{Non-linear superfields from linear ones}\label{ss:IvanovKapustnikov}

In two seminal papers, Callan, Coleman, Wess and Zumino discussed phenomenological Lagrangians invariant under non-linear realizations of a broken symmetry group \cite{Coleman:1969sm,Callan:1969sn}. We can summarize their result as \emph{any linear multiplet of a given group can be converted into the direct sum of non-linearly transforming fields, by means of the group transformation with the Goldstone field as a parameter}. The analogous theorem also holds for supersymmetry \cite{Ivanov:1977my,Ivanov:1978mx,Ivanov:1982bpa}.

We choose the superspace notation in four dimensions of \cite{Wess:1992cp}, for four-component spinors, see \cite{Ivanov:1977my,Ivanov:1978mx}. Take any superfield $\Phi(x,\theta,\bar \theta)$. Recall that supersymmetry acts on a superfield as
\begin{equation}
\Phi'(x,\theta,\bar \theta) = \exp(\xi Q +\bar \xi  \bar Q) \Phi(x,\theta,\bar \theta) = \Phi(x+\rmi \theta \sigma \bar \xi - \rmi  \xi \sigma \bar \theta,\theta+\xi,\bar \theta+\bar \xi)\,,\label{eq:nonlinear_general}
\end{equation}
with the supersymmetry generators
\begin{equation}
 Q_\alpha = \frac{\partial}{\partial \theta^\alpha} - \rmi \sigma^\mu_{\alpha \dot \alpha} \bar \theta^{\dot \alpha}\partial_\mu\,,\qquad \bar Q^{\dot \alpha} = \frac {\partial}{\partial \theta_{\dot \alpha}} - \rmi \theta^\alpha \sigma^\mu_{\alpha \dot \beta} \epsilon^{\dot \beta \dot \alpha}\,.
\end{equation}
If supersymnmetry is spontaneously broken, there is a Goldstino $\zeta$, which transforms under the broken SUSY generators as 
\begin{equation}
 \delta_\xi \gold (x) = \xi - \rmi   v^\mu \partial_\mu \gold(x)\,,\qquad v^\mu =  \gold(x) \sigma^\mu \bar \xi - \xi \sigma^\mu \bar \gold(x)\,.
\end{equation}
Consider the superfield $\hat \Phi$ defined by acting on $\Phi$ with the broken SUSY transformation with  the Goldstino as group parameter $\xi \to -\gold$:
\ba
  \hat \Phi(x,\theta,\bar \theta) &\equiv&  \Phi(x',\theta',\bar \theta') \,,
\ea
with $x'= x-\rmi \theta \sigma \bar \gold + \rmi  \gold \sigma \bar \theta$, $\theta'=\theta-\gold$ and $\bar \theta'=\bar \theta-\bar \gold$.
Then $\hat \Phi$ transforms non-linearly as
\begin{equation} 
\delta_\xi \hat \Phi(x,\theta,\bar \theta) = -\rmi   v^\mu \hat \Phi (x,\theta,\bar \theta)\,.\label{eq:nonchiral_transformation}
\end{equation}
In particular, \emph{all components of $\hat \Phi$ transform individually in the same way!} One concludes that the standard non-linear transformation on any matter field $\phi$ (scalar, spinor, vector,\ldots) is
\begin{equation}
 \delta_\xi \phi =  -\rmi  (\gold \sigma^\mu \bar \xi(x) - \xi(x) \sigma^\mu \bar \gold) \partial_\mu \phi\,, \label{eq:nonLinearTransfo}
\end{equation}
or in Majorana spinor notation $\epsilon \equiv(\xi,\bar \xi), \lambda \equiv (\gold,\bar \gold)$ 
\begin{equation}
 \delta_\epsilon \phi = \bar \lambda \gamma^\mu \epsilon \partial_\mu \phi\,.
\end{equation}

\subsection{Chiral notation}\label{ss:chiral}

For chiral superfields it is convenient to consider a redefinition such that the Goldstino $\gold$ does not mix with $\goldbar$. Introduce a spinor $\tgold$ that transforms only into itself \cite{Zumino:1974qc,Samuel:1982uh}:
\begin{align} \label{eq:Goldstino_chiral}
\delta_\xi \tgold_\alpha =  \xi_\alpha - 2 \rmi  \kap\tgold \sigma^m \bar \xi \partial_m \tgold_\alpha\,,\qquad 
\delta_\xi \bar \tgold_{\dot \alpha} &= \bar \xi_{\dot \alpha} +2 \rmi \kap \xi \sigma^m \bar \tgold \partial_m \bar \tgold _{\dot \alpha} \,.
\end{align}
The two spinors are related as
\begin{align}
 \tgold_\alpha  (x) &=  \gold(z)\,,\qquad \bar \tgold_{\dot \alpha}  (x) =  \bar {\gold} (z^*)\,,\label{eq:def_tilde_lambda}
\end{align}
with 
\begin{equation}
  z^m = x^m - \rmi \gold (z) \sigma^m \bar {\gold} (z)\,, \qquad
 (z^*)^m = x^m + \rmi \gold(z^*) \sigma^m \bar {\gold} (z^*)\,.
\end{equation}
One can solve this in a Taylor series, which terminates because of the anti-commuting spinors  \cite{Samuel:1982uh}:
\ba
\tgold _\alpha &=& \gold _\alpha - \rmi  \tilde v^m \partial_m \gold _\alpha  - \frac 12  \tilde v^m \tilde v^n \partial_m \partial_n \gold_\alpha -  \tilde v^m (\partial_m \tilde v^n) \partial_n \gold_\alpha\cr
&& +\rmi   \tilde v^\ell (\partial_\ell \tilde v^m) (\partial_m \tilde v^n) \partial_n \gold _\alpha
+ \frac 12 \rmi   v^\ell v^m (\partial_\ell \partial_m v^n) \partial_n \gold_\alpha\,,\label{eq:Goldstino_chiral_relation}
\ea
with $\tilde v^m \equiv \gold \sigma^m \bar {\gold}$.

The same implicit redefinition through a shift of the coordinate $x\to z$ also works for other fields:
\begin{equation}
 \delta_\xi \tilde\phi =- 2 \rmi  \kap\tgold \sigma^m \bar \xi \partial_m\tilde \phi\,,\label{eq:chiral_transformation}
\end{equation}
with
\begin{equation}
 \tilde \phi(x) = \phi(z)\,.\label{eq:tildenotation}
\end{equation}

\subsection{Example: chiral superfield}

Let's see explicitly how a chiral superfield can be turned into a non-linearly transforming one. Consider an unconstrained chiral superfield $\Phi$
\begin{equation}
 \Phi(x,\theta,\bar \theta) = \phi(y) + \sqrt{2}  \psi(y) \theta + F_\phi(y) \theta^2\,,\qquad y= x - \rmi  \theta \sigma \bar \theta\,.
\end{equation}
Following the general prescription \eqref{eq:nonlinear_general}, the non-linearly transforming superfield is
\begin{equation}
 \hat \Phi(x,\theta,\bar\theta) = \phi(y') + \sqrt{2} \psi (y') (\theta - \gold(x)) + F_\phi (y') (\theta - \gold(x))^2\,,
\end{equation}
where the $x$-dependence is implicit through $y' = x' -i \theta' \sigma \bar \theta'$ or
\begin{equation}
 y' = y - 2 \rmi  \theta  \sigma^\mu \bar \gold(x) + \rmi  \gold(x)  \sigma \bar \gold(x)\,.
\end{equation}
We check this explicitly in components. We can define new components as
\begin{equation}
 \hat \Phi(x,\theta,\bar\theta) \equiv \hat \phi(x) + \sqrt 2 \hat \psi (x) \theta + \hat F_\phi(x) \theta^2 + \text{terms with $\bar\theta, \bar \theta^2$}
\end{equation}
Each of these components transforms non-linearly as in \eqref{eq:nonLinearTransfo}. Its lowest two components are
\begin{align}\label{eq:nonlinear_chiral}
 \hat  \phi (x) &\equiv \phi(\hat x) - (\sqrt 2 \psi (\hat x)\gold(x) -F_\phi(\hat x) \gold^2(x))\,,\\
 \hat \psi(x) &\equiv \psi(\hat x) - \sqrt 2 F_\phi(\hat x) \gold (x) - \sqrt 2 \rmi  (\sigma^\mu \bar \gold(x)) \left[\partial_\mu \phi(\hat x) - \sqrt2 \gold(x) \partial_\mu \psi(\hat x) + \gold^2(x)\partial_\mu F_\phi(\hat x) \right]\,,\nn
\end{align}
where we also introduced the notation
\begin{equation}
 \hat x = x + \rmi  \gold(x) \sigma \bar \gold(x).
\end{equation}
One can make the right-hand sides of \eqref{eq:nonlinear_chiral} dependent on $\hat x$ only, by using $\gold (x) = \tgold (\hat x)$ from \eqref{eq:def_tilde_lambda}.

\subsection{Superfield constraints to remove components}

We have seen that we do not need to constrain superfields to find non-linear transformations. Constraints are however useful tools to eliminate certain degrees of freedom. One can see that the constraints of \cite{Komargodski:2009rz} exactly eliminate one or several components of the non-linearly transforming fields. For an all-encompassing view of different constraint multiplets in  supergravity see \cite{DallAgata:2016yof}.

\subsubsection{Nilpotent chiral superfield: Goldstino}

A nilpotent chiral superfield $S = s + \sqrt 2 G \theta + F \theta^2$  can describe a goldstino. We derive how the nilpotency constraint $S^2=0$ leads to the chiral goldstino $\tgold = G/2F$, or with the redefinition \eqref{eq:def_tilde_lambda}
\begin{equation}
 \gold(x) = \frac{G(z) }{2 F(z)}\,,\qquad z^m = x^m - \rmi  \gold (z) \sigma^m \bar {\gold} (z)\,.
\end{equation}

Consider first an unconstrained chiral superfield $S$, with supersymmetry transformations
\begin{align}
 \delta_\xi s &= \sqrt 2 \xi G\,,\nonumber\\
 \delta_\xi G_\alpha &= \sqrt 2 F \xi_\alpha + \rmi  \sqrt 2 (\sigma^\mu \bar \xi)_\alpha \partial_\mu s\,,\nonumber\\
 \delta_\xi F &= \rmi  \sqrt 2\bar \xi \bar \sigma ^\mu \partial_\mu G\,.
 \label{eq_susy_KS}
\end{align}
Then the supersymmetry transformation of $\tgold= \frac{G}{\sqrt 2 F}$ is exactly
\begin{align}
\delta_\xi  \tgold_\alpha &= \frac{\delta_\xi G_\alpha }{\sqrt 2 F} - \frac {G_{\alpha}} {\sqrt 2 F^ 2}\delta_\xi F\\
&= \xi_\alpha -2 \rmi  (\tgold \sigma^\mu  \bar  \xi )\, \partial_\mu \tgold_\alpha - \frac \rmi  F (\sigma^\mu \bar \xi)_\alpha \partial_\mu  \left( s - \frac{G^2}{2F}\right)\,.\label{eq:afterFierz}
\end{align}
We see that the identification $s = G^2/2F$, which follows from $S^2 =0$, indeed gives the standard non-linear transformation of the Goldstino.

\subsubsection{Orthogonal chiral superfield: fermion}

Take a chiral superfield $\Phi$ and enforce the additional constraint, giving ``orthogonal'' superfields (a terminology proposed in \cite{Ferrara:2015tyn})
\begin{equation}
 S \Phi = 0\,.
\end{equation}
As explained in \cite{Komargodski:2009rz}, this gives the relation for the scalar component
\begin{equation}
 \phi = \frac{\psi G}{F} - \frac{G^2}{2 F^2} F_\phi = \sqrt 2 \psi_q \tgold- \tgold^2F_q\,.
\end{equation}
Comparing to \eqref{eq:nonlinear_chiral}, and using $\tgold(\hat x) = \gold (x)$, we see that this elimates the non-linearly transforming scalar component $\hat \phi = 0$. Hence we are only left with the non-linearly transforming fermion $\hat \psi$, whose transformation becomes
\begin{align}
 \hat \psi(x) &= \psi(\hat x) - \sqrt 2 F(\hat x) \gold (x) - \sqrt 2 \rmi  (\sigma^\mu {\bar \gold}(x) )\left[\sqrt2 \psi(\hat x)  \partial_\mu \gold(x)  -2 F(\hat x) \gold(x) \partial_\mu \gold(x)\right]\,.
\end{align}

\subsubsection{Other constraints}

Other constraints that are used in a cosmology setting, are constraints that only keep bosonic degrees of freedom in a superfield. For a chiral superfield $H$,  the constraint $S \bar D_{\dot \alpha} \overline{\cal H}_{NL} =0$ eliminates the fermion and leaves only a complex boson. For a field-strength superfield $W_\alpha$, describing a vector field and a gaugino, the constraint $S W_\alpha =0$ removes the gaugino. One can study other constraints, for an all-encompassing view see \cite{DallAgata:2016yof}.

\bibliographystyle{JHEP}
\bibliography{refs}

\end{document}